\documentclass[twocolumn,showpacs,preprintnumbers,amsmath,amssymb,showkeys]{revtex4}
\usepackage{graphicx}
\usepackage{dcolumn}
\usepackage{bm}
\newcommand{\be}{\begin{equation}}
\newcommand{\ee}{\end{equation}}
\newcommand{\ba}{\begin{eqnarray}}
\newcommand{\ea}{\end{eqnarray}}

\begin{document}

\title{Properties of new unflavored mesons below 2.4 GeV}

\author{S. S. Afonin}
\affiliation{V. A. Fock Department of Theoretical Physics, St. Petersburg State University,
1 ul. Ulyanovskaya, 198504, St. Petersburg, Russia.}
\email{afonin24@mail.ru}


\begin{abstract}
The global features of spectrum of highly excited light
nonstrange mesons can be well understood within both chiral
symmetry restoration scenario combined with the relation
$M^2\sim J+n$ and within nonrelativistic
description based on the relation $M^2\sim L+n$. The predictions
of these two alternative classifications for missing states are
different and only future experiments can distinguish between the two.
We elaborate and compare systematically the predictions of both schemes, which
may serve as a suggestion for future experiments devoted to the search
for missing states.
\end{abstract}

\maketitle

\section{INTRODUCTION}

In recent years many new data on unflavored mesons have appeared in
the section "Further States" of Particle Data Group~\cite{pdg}. The main
source for these data came from the Crystal Barrel experiment,
where plenty of new states were observed in the proton-antiproton
annihilation in the energy range 1.9-2.4~GeV~\cite{ani,bugg}. The obtained
spectrum remarkably confirmed the approximate linearity of both Regge trajectories
and radial Regge trajectories (or, equivalently, the equidistance of daughter
trajectories). An important feature of the spectrum is that the slopes
of both types of trajectories are almost equal, i.e., the following
relation can be written (see, e.g.,~\cite{a1,a2} for discussions):
\be
\label{01}
M^2_i\sim J+n+c_i,
\ee
where $i$ denotes a set of quantum numbers, $J$ is the spin, $n$ is the "radial"
quantum number, and $c_i$ is a constant. Theoretically such type of mass formulas
appeared in dual~\cite{gLS}, hadron string~\cite{zw},
and AdS/QCD~\cite{katz} models. The experimental spectrum of unflavored mesons
reveals a clear-cut clustering of states near certain equidistant values
of masses square~\cite{a2}, which implies that the constants $c_i$ should
be equal or differ by an integer. If we fit the experimental data
by means of Eq.~\eqref{01}, the constants $c_i$ will not be universal and
a relation between different $c_i$ {\it a priori} is not clear.

However, instead of Eq.~\eqref{01}, one can consider its nonrelativistic
analog~\cite{sh,a4,a5,glozrev},
\be
\label{1}
M^2_i\sim L+n+c,
\ee
with the angular momentum of quark-antiquark pair $L$ being related to the total
spin $J$ as $J=L,\,L\pm1$ depending on the mutual orientation of the quark/antiquark
spin $s$. It turns out that the angular momentum assignment can be chosen
such that the constant $c$ will be approximately universal, as is
written in Eq.~\eqref{1}.
This means, in particular, that $L$ and quark spins $s$ can be added as
in the usual quantum mechanics.
Such a physical picture is quite unexpected because light mesons are
ultrarelativistic systems, therefore $L$ and $s$ cannot be separated,
a conserved quantum number is the total spin $J$, while $L$ would be
conserved with the spinless quarks only. The validity of Eq.~\eqref{1}
could be a nontrivial consequence of the asymptotic suppression of
the spin-orbital correlations in excited hadrons~\cite{sh,glozrev,wil,matag}.

Relation~\eqref{1} implies a duplication of states
in the channels where the resonances can be created by different angular momentum.
For instance, the vector mesons can have either $L=0$ or $L=2$ (the so-called S- and D-wave
mesons in the quantum-mechanical terminology), hence, they are duplicated. Experimentally such a
duplication is well seen~\cite{ani,bugg}. In practice, the separation of resonances into
the states with different angular momentum can be achieved by using the polarization data.
Following this method, the experiment of the Crystal Barrel Collaboration obtained a good separation
for the states with $(C,I)=(+1,0),(-1,1)$~\cite{bugg}. The separation in other channels
should be tentatively guessed. As long as one accepts a nonrelativistic framework, the parity
of quark-antiquark pair is defined as $P=(-1)^{L+1}$. The states with maximal $L$ at given mass
are then parity singlets, associating them with the resonances on the leading
Regge trajectories, we obtain a correct qualitative picture of the known
experimental spectrum.

Another pattern of parity doubling is predicted by the chiral symmetry
restoration (CSR) scenario (see~\cite{glozrev} for a review). If effective
CSR occurs high in the spectrum, the chiral multiplets become complete.
In particular, this implies the absence of parity singlet states among
highly excited hadrons. Within the CSR picture, the duplication of some
trajectories appears due to an assignment of states on these trajectories
to different chiral multiplets.

The classifications of states based on CSR and the ones based on
Eq.~\eqref{1} cannot coexist because the relativistic chiral basis and
the nonrelativistic $n^{2s+1}\!L_{J}$ basis are incompatible~\cite{gln},
the chiral basis, however, can meet Eq.~\eqref{01}.

Thus, an intriguing problem emerges --- which alternative (if any) is
realized in nature? The answer can be provided by examining the
phenomenological implications of the possibilities above, such as
spectroscopic predictions. A phenomenological analysis of these
predictions is still absent in the literature and the present
paper is intended to fill in this gap, providing thereby a
stimulus for the search of new states that distinguish between the two
alternatives.

We will show by an explicit assignment of mesons according to the quantum numbers $(L,n)$ that
relation~\eqref{1} describes the spectrum of practically all confirmed and unconfirmed unflavored
mesons except the masses of Goldstone bosons. There are only eight missing
states below 2.4 GeV, which allow to justify or falsify the classification in future.
The CSR scenario predicts these eight states as well, but it predicts also many missing
states beyond them.

The paper is organized as follows. In Sec.~II we remind the reader of some phenomenological
ideas concerning the origin of linear spectrum and estimate qualitatively
an expected value for the constant $c$ in Eq.~\eqref{1}.
Section~III contains our phenomenological
analysis and predictions. We conclude in Sec.~IV.

\section{THEORETICAL DISCUSSIONS}

Let us present some known heuristic arguments in favor of
linear spectrum. For high radial or orbital excitation, a meson
state can be considered quasiclassically as a pair of relativistic
quarks interacting via a linear potential. Consequently,
neglecting the quark spin, the meson mass can be written as
\be
\label{2} M=2p+\sigma r,
\ee
where $p$ is the relativistic quark
momentum and $\sigma$ is the string tension. The maximal length of
the chromoelectric flux tube between the quarks is $l=M/\sigma$.
Applying the quasiclassical (WKB) quantization condition,
\be
\label{3} \int_0^lp\,dr=\pi n,
\ee
with the momentum $p$ taken from Eq.~\eqref{2}, one obtains
\be
\label{4}
M^2\sim n.
\ee

A "next-to-leading" correction to the presented picture can be considered. It comes from the
Bohr-Sommerfeld quantization condition~\eqref{3}: $n$ must be replaced by $n+\gamma$, where
$\gamma$ is a constant of order of unity characterizing the nature of turning points. In Eq.~\eqref{2}
one deals with a centrosymmetrical potential. It is well known (see, e.g.,~\cite{landau}) that
in this case $\gamma=\frac12$. Hence, the corrected linear spectrum is
\be
\label{5}
M^2\sim n+\frac12.
\ee
Exactly this type of spectrum is predicted by the Lovelace-Shapiro dual
amplitude~\cite{LS}, where $\gamma=\frac12$ comes from the Adler
self-consistency condition (at $p^2=m_{\pi}^2$, the $\pi\pi$
scattering amplitude is zero). In some channels this spectrum appeared
naturally within the QCD sum rules~\cite{a3}, where $\gamma=\frac12$ stems
from the absence of dimension-two gauge-invariant condensate.
Recently the intercept $\frac12$ has been reported within a
holographic dual of QCD (the second reference in~\cite{katz}).

Specific boundary conditions can lead to another value for $\gamma$. We
mention the following possibilities: identified ends (closed string)
correspond to $\gamma=0$, $S$-wave states correspond to $\gamma=\frac34$,
infinite potential walls at the ends correspond to $\gamma=1$. The first
possibility is unrealistic for mesons, thus in a general case we expect
$\gamma$ to lie in the interval $\frac12\leq\gamma\leq1$.

According to Regge theory and simple hadron string considerations,
$M^2$ is also linear in the angular
momentum $L$ (Chew-Frautschi formula). This suggest that
$n$ in Eq.~\eqref{4} might be substituted by $n+L$, thus,
resulting in Eq.~\eqref{1}. Unfortunately, we are not aware of
solid arguments for such a replacement.

The linear spectrum~\eqref{4} is an exact result within a kind of
dimension-two QCD, the 't Hooft model~\cite{dim2}. The next-to-leading correction
to Eq.~\eqref{4} within this model, however, is $O(\ln{n})$ rather
than a constant. In this respect we should remind the reader that the 't Hooft model is defined in
a specific sequence of $N_c\rightarrow\infty$ limits, $m_q\rightarrow0$
while $m_q\gg g\sim1/\sqrt{N_c}$, where $m_q$ denotes current quark mass and $g$ is
coupling constant. In contrast to QCD, we cannot set $m_q=0$ from the very beginning.
On the other hand, if one takes into
account the masses of current quarks in the derivation above, the
logarithmic corrections emerge naturally (see, e.g.,~\cite{simonov}).

A delicate point in such kind of reasoning is the relative value of
slope between radial and orbital trajectories. The matter is that
$M^2=4\pi\sigma$ in the derivation above, but $M^2=2\pi\sigma$
according to the Chew-Frautschi formula. Naively, this leads to
$M^2\sim L+2n$ rather than to Eq.~\eqref{1}. A possible reason is
that parity is not properly incorporated: It is related to the
orbital motion (defined through $L$) in three space dimensions,
but in one space dimension it is related to the reflections of
wave functions. Considering the radial excitations of a
one-dimensional object, one deals with the latter case, where the
states alternate in parity, like in the 't Hooft model. The extraction
of states with the same parity is then tantamount to enlarging of
the slope by two times.

The note above is a particular manifestation of a general problem:
A linear potential plus a semiclassical analysis produces a necessarily
different angular and radial slopes, for this reason it may be suggestive
only and by no means may serve for justification of Eq.~\eqref{1}.
A derivation of Eq.~\eqref{1} or Eq.~\eqref{01} is a challenge for future
quark models~\cite{bicudo}, presently these empirical relations do not
have solid theoretical support. In particular, Eq.~\eqref{1} implies
the existence of a single "principal" quantum number, $N=L+n$, like
in a hydrogen atom~\cite{a5}, a development of this analogy
could be far reaching.

\section{FITS AND PREDICTIONS}

Using experimental masses from the Particle Data Group~\cite{pdg} one can perform a global fit
of the data by the linear spectrum. Such an analysis was performed
in~\cite{a2}. The result is that on average the masses of well known
light nonstrange mesons behave as (in GeV$^2$)
\be
\label{6}
M^2_{\text{exp}}\approx1.14(N+0.54), \qquad N=0,1,2.
\ee
One can consider the states observed by the Crystal Barrel
experiment~\cite{bugg}, which allow us to extend Eq.~\eqref{6} to
$N=3,4$. It turns out that both slope and intercept are then changed
negligibly~\cite{a2}. Comparing Eqs.~\eqref{5} and~\eqref{6}
we see that our guess on the "next-to-leading" correction
is well compatible with the experimental data.

Partly following~\cite{ani,bugg}, we classify the light
nonstrange mesons according
to the values of $(L,n)$, see Table~\ref{t1}. As seen from Table~\ref{t1},
the states with equal $N=L+n$ are indeed approximately degenerate
(one should read the data in a diagonal way, the frames are
introduced for convenience).
We will regard the averaged values of masses and widths at given
$N$ from~\cite{a2} as predictions for unknown states in the mass region under
consideration. Thus, for $M(N)$ we have (in MeV):
$M(0)\approx 785$, $M(1)\approx 1325\pm90$, $M(2)\approx 1700\pm60$
$M(3)\approx 2000\pm40$, $M(4)\approx 2270\pm 40$.
Looking at Table~\ref{t1}, we make the following predictions for the
nonstrange mesons which still have not been observed.
\begin{enumerate}
  \item In the energy range $1700\pm60$ MeV there exists $a_0$, $f_1$, $\rho_2$,
  $\omega_2$, as well as the second $\rho$ and $\omega$ mesons.
  Their widths are approximately $\Gamma=200\pm70$ MeV. The state $X(1650)$ with
  $I^G(J^{PC})=0^-(?^{?-})$
  cited in~\cite{pdg} might be a possible candidate for the predicted
  $\omega$ or $\omega_2$ mesons. The state $X(1750)$ with $I^G(J^{PC})=?^?(1^{--})$
  cited in~\cite{pdg} might be a possible candidate for the predicted
  $\omega$ or $\rho$ mesons.
  \item In the energy range $2000\pm40$ MeV there exists the second $\omega$ meson.
  Its width is approximately $\Gamma=220\pm70$ MeV. The state $X(1975)$ with
  $I^G(J^{PC})=?^?(?^{??})$ cited in~\cite{pdg} might be a possible candidate for the
  predicted $\omega$ meson.
  \item In the energy range $2270\pm40$ MeV there exists $a_0$ meson.
  Its width is approximately $\Gamma=270\pm60$ MeV. The states $X(2210)$ and $X(2340)$
  with $I^G(J^{PC})=?^?(?^{??})$ cited in~\cite{pdg} might be possible candidates for the
  predicted $a_0$ meson.
\end{enumerate}

\begin{table*}
\caption{\label{t1} Classification of light nonstrange mesons according
to the values of $(L,n)$. The states with the lowest star rating (according to~\cite{bugg})
are marked by the question mark, the states, which presumably have a large admixture
of strange quark, are marked by the double question mark.}
\begin{ruledtabular}
\begin{tabular}{cccccc}
\begin{tabular}{c}
\begin{picture}(15,15)
\put(0,15){\line(1,-1){15}}
\put(0,0){$L$}
\put(10,10){$n$}
\end{picture}\\
\end{tabular}
& 0 & 1 & 2 & 3 & 4 \\
\hline
0
&
\begin{tabular}{c}
$\pi(140)$\\
$\eta(548)$(??)\\
$\rho(770)$\\
$\omega(782)$\\
\end{tabular}
&
\begin{tabular}{|c|}
\hline
$\pi(1300)$\\
$\eta(1295)(??)$\\
$\rho(1450)$\\
$\omega(1420)$\\
\hline
\end{tabular}
&
\begin{tabular}{c}
$\pi(1800)$\\
$\eta(1760)$\\
$\rho(?)$\\
$\omega(?)$\\
\end{tabular}
&
\begin{tabular}{||c||}
\hline
$\pi(2070)$\\
$\eta(2010)$\\
$\rho(1900)$\\
$\omega(?)$\\
\hline
\end{tabular}
&
\begin{tabular}{c}
$\pi(2360)$\\
$\eta(2285)$\\
$\rho(2150)$\\
$\omega(2205)$(?)\\
\end{tabular}
\\
1
&
\begin{tabular}{|c|}
\hline
$f_0(1370)$\\
$a_0(1450)$(??)\\
$a_1(1260)$\\
$f_1(1285)$\\
$b_1(1230)$\\
$h_1(1170)$\\
$a_2(1320)$\\
$f_2(1275)$\\
\hline
\end{tabular}
&
\begin{tabular}{c}
$f_0(1770)$\\
$a_0(?)$\\
$a_1(1640)$\\
$f_1(?)$\\
$b_1(1620)$(?)\\
$h_1(1595)$(?)\\
$a_2(1680)$\\
$f_2(1640)$\\
\end{tabular}
&
\begin{tabular}{||c||}
\hline
$f_0(2020)$\\
$a_0(2025)$\\
$a_1(1930)$(?)\\
$f_1(1971)$\\
$b_1(1960)$\\
$h_1(1965)$\\
$a_2(1950)$(?)\\
$f_2(1934)$\\
\hline
\end{tabular}
&
\begin{tabular}{c}
$f_0(2337)$\\
$a_0(?)$\\
$a_1(2270)$(?)\\
$f_1(2310)$\\
$b_1(2240)$\\
$h_1(2215)$\\
$a_2(2175)$(?)\\
$f_2(2240)$\\
\end{tabular}
&\\
2
&
\begin{tabular}{c}
$\rho(1700)$\\
$\omega(1650)$\\
$\pi_2(1670)$\\
$\eta_2(1645)$\\
$\rho_2(?)$\\
$\omega_2(?)$\\
$\rho_3(1690)$\\
$\omega_3(1670)$\\
\end{tabular}
&
\begin{tabular}{||c||}
\hline
$\rho(2000)$\\
$\omega(1960)$\\
$\pi_2(2005)$\\
$\eta_2(2030)$\\
$\rho_2(1940)$\\
$\omega_2(1975)$\\
$\rho_3(1982)$\\
$\omega_3(1945)$\\
\hline
\end{tabular}
&
\begin{tabular}{c}
$\rho(2265)$\\
$\omega(2295)$(?)\\
$\pi_2(2245)$\\
$\eta_2(2267)$\\
$\rho_2(2225)$\\
$\omega_2(2195)$\\
$\rho_3(2300)$(?)\\
$\omega_3(2285)$\\
\end{tabular}
&  &\\
3
&
\begin{tabular}{||c||}
\hline
$f_2(2001)$\\
$a_2(2030)$\\
$f_3(2048)$\\
$a_3(2031)$\\
$b_3(2032)$\\
$h_3(2025)$\\
$f_4(2018)$\\
$a_4(2005)$\\
\hline
\end{tabular}
&
\begin{tabular}{c}
$f_2(2293)$\\
$a_2(2255)$\\
$f_3(2303)$\\
$a_3(2275)$\\
$b_3(2245)$\\
$h_3(2275)$\\
$f_4(2283)$\\
$a_4(2255)$\\
\end{tabular}
&  &  &\\
4
&
\begin{tabular}{c}
$\rho_3(2260)$\\
$\omega_3(2255)$\\
$\rho_4(2230)$\\
$\omega_4(2250)$(?)\\
$\pi_4(2250)$\\
$\eta_4(2328)$\\
$\rho_5(2300)$\\
$\omega_5(2250)$\\
\end{tabular}
&  &  &  &\\
\end{tabular}
\end{ruledtabular}
\end{table*}

Thus, the nonrelativistic $n^{2s+1}\!L_J$ assignment based
on Eq.~\eqref{1} predicts eight nonstrange mesons
in the energy range 1.6-2.3 GeV which have never been observed and are
awaiting their discovery.

Consider predictions of the CSR scenario based on Eq.~\eqref{01}. Evidently,
all eight missing states above should also follow from this scenario if
effective CSR takes place above 1.7 GeV. We will enumerate the predictions
which go beyond these eight new mesons.

\begin{enumerate}
  \item $1700\pm60$ MeV. The indications on CSR are not solid in this mass
  region. Nevertheless, if CSR happens we may expect in the minimal scenario
  the appearance of parity partners for $\rho_3$ and $\omega_3$ mesons ---
  new $a_3$ and $f_3$ mesons, respectively. If CSR leads to parity-chiral
  multiplets described in~\cite{glozrev} [the $(1,0)\oplus(0,1)$ and
  $(\frac12,\frac12)$ representations of $SU(2)_L\times SU(2)_R$)] then we
  should expect also the second
  $\rho_3$ and $\omega_3$ mesons and their $(\frac12,\frac12)$ chiral
  partners, the $h_3$ and $b_3$ mesons.
  \item $2000\pm40$ MeV. We should expect at least the parity partners
  for $a_4$ and $f_4$ mesons --- the states $\rho_4$ and $\omega_4$.
  If CSR results in parity-chiral
  multiplets described in~\cite{glozrev} then we should expect also
  the second $a_4$ and $f_4$ states, their chiral partners $\eta_4$ and $\pi_4$,
  and the second $\rho_3$ and $\omega_3$ mesons [all carry the representation
  $(\frac12,\frac12)$].
  \item $2270\pm40$ MeV. We should expect at least the parity partners
  for $\rho_5$ and $\omega_5$ mesons --- the states $a_5$ and $f_5$.
  If CSR leads to parity-chiral
  multiplets described in~\cite{glozrev} then we should expect
  also the second $\rho_5$ and $\omega_5$ states, their
  chiral partners $h_5$ and $b_5$, and the
  second $a_4$ and $f_4$ mesons [all carry the
  representation $(\frac12,\frac12)$].
\end{enumerate}

Thus, the CSR scenario combined with a clustering of states expressed by
Eq.~\eqref{01} leads to a richer spectrum of high excitations.

\section{CONCLUSIONS}

We have provided in a concise form the concrete spectroscopic predictions
which follow from recent discussions on global features of a light
nonstrange meson spectrum.

The assumption that relation~\eqref{1} does not depend on quantum numbers of unflavored nonexotic
mesons allows us to provide the whole spectrum with two input parameters only, the universal slope and
intercept. The quasiclassical and some other arguments indicate that these inputs could be related.
Fixing the physical values for the slope and
intercept, universal relation~\eqref{1} gives 100 nonstrange mesons below 2.4~GeV, see Table~\ref{t1}.
Except in some rare cases, e.g., the Goldstone bosons, the agreement with the masses of known confirmed
resonances from the Particle Data Group~\cite{pdg} and unconfirmed states observed by Crystal Barrel~\cite{bugg}
is impressive. There exist only eight missing states which have never been observed. The predictions
for their masses and widths are given and possible candidates are indicated. We do not see any
theoretical reasons why those states should be absent in nature, most likely they still have been not
detected experimentally. The seemingly random (factor isospin) distribution of missing states on the spectrum
supports our belief.

Relation~\eqref{1} is at odds with the Lorentz group (angular momentum $L$
is not conserved quantum number in relativistic quark-antiquark pair)
and chiral symmetry restoration. Both obstacles can be overcome if
one accepts relation~\eqref{01}, the number of predicted states below
2.4 GeV is then substantially larger.

The discovery of indicated missing resonances in future experiments will
constitute a crucial test for the two alternatives discussed in the paper,
providing thereby an important step forward toward establishing final
order in the spectroscopy of light mesons.

\section*{ACKNOWLEDGMENTS}
The work was supported by RFBR, grant no. 05-02-17477, by the
Ministry of Education of Russion Federation, grant no. RNP.2.1.1.1112,
and by grant no. LSS-5538.2006.2.


\end{document}